\documentclass{article}
\usepackage{graphicx}
\usepackage{cite}
\begin{document}

\title{\bf Optimized basis expansion as an extremely accurate
technique for solving time-independent Schr\"{o}dinger equation}
\author{Pouria Pedram\thanks{Email: p.pedram@srbiau.ac.ir},$^a$ Mahdi Mirzaei,$^b$ and S.S. Gousheh$^b$
\\  {\small $^a$Department of Physics, Science and Research Branch, Islamic
        Azad University, Tehran, Iran}\\
        {\small $^b$Department of Physics, Shahid Beheshti University,
Evin, Tehran 19839, Iran}}

\maketitle \baselineskip 24pt

\begin{abstract}
We use the optimized trigonometric finite basis method to find
energy eigenvalues and eigenfunctions of the time-independent
Schr\"{o}dinger equation with high accuracy. We apply this method to
the quartic anharmonic oscillator and the harmonic oscillator
perturbed by a trigonometric anharmonic term as not exactly solvable
cases and obtain the nearly exact solutions.

\vspace{5mm}
{\it Keywords:} Schr\"{o}dinger equation; Anharmonic oscillator; Finite basis method; Variational scheme
\end{abstract}
\maketitle
% ----------------------------------------------------------------
\section{Introduction}
Eighty years after the birth of quantum mechanics, the
Schr\"{o}dinger's famous equation still remains a subject for
numerous studies, aiming at extending its field of applications and
at developing more efficient analytic and approximation methods for
obtaining its solutions. There has always been a remarkable interest
in studying exactly solvable Schr\"{o}dinger equations. In this
sense, the exact solubility has been found for only a very limited
number of potentials, most of them being classified already by
Infeld and Hull \cite{Infeld} on the basis of the Schr\"{o}dinger
factorization method \cite{factorization,p1}. However, a vast majority
of the problems of physical interest do not fall in the above
category and we have to resort to approximation techniques.

The need for such methods have stimulated development of more sophisticated
integration approaches such as embedded exponentially-fitted
Runge-Kutta \cite{Runge-Kutta}, dissipative Numerov-type
method \cite{Numerov},
relaxational approach \cite{Relax} based on the Henyey algorithm
\cite{Henyey}, an adaptive basis set using a hierarchical finite
element method \cite{hierarchical}, and an approach based on
microgenetic algorithm \cite{microgenetic}, which is a variation of
a global optimization strategy proposed by Holland \cite{Holland}.
We can also mention the variational
sinc collocation method \cite{Amore} and the Instanton
method \cite{Ulrich}.

In this paper, we expand the wave function in terms of an
orthonormal set of the eigenfunctions of a Hermitian operator,
namely, the basis-set expansion method. Indeed, we use the
trigonometric basis functions obeying periodic boundary condition.
The accuracy of the solutions strongly depends on the domain of the
wave function. So we implement the Rayleigh-Ritz variational method
to find the domain's optimal value. The application  of this method
for the Dirichlet boundary condition is also discussed in
Ref.~\cite{taseli,Bhattacharyya}. Note that, an analytic relation
for this optimal length is recently presented in Ref.~\cite{p4}
which is only applicable for the power low potentials. A
two-dimensional application of this method is also discussed in the
context of quantum cosmology \cite{p3}.

The remainder of this paper is organized as follows. In Sec.~2, we
present the underlying theoretical bases for the formulation of the
trigonometric basis-set expansion method and the optimization
procedures. In Sec.~3, to illustrate the method, we apply this
method for the Simple Harmonic Oscillator. We then solve two
perturbed harmonic oscillators that are not exactly solvable, the
first with a quartic anharmonic term, and second with a rapidly
oscillating trigonometric anharmonic term. We present our
conclusions in Sec.~4.

\section{The trigonometric basis-set expansion method}
Let us consider the time-independent one-dimensional Schr\"{o}dinger
equation
\begin{equation}\label{Schrodinger}
-\frac{\hbar^2}{2m}\frac{d^2\psi(x)}{dx^2}+V(x)\psi(x)=E\psi(x),
\end{equation}
where $m$, $V(x)$, and $E$ stand for the reduced mass, potential, and energy, respectively. This equation can be written in
the form
\begin{equation}\label{ODE}
-\frac{d^2\psi(x)}{dx^2}+\hat{f}(x)\psi(x)=\epsilon\,\psi(x),
\end{equation}
where
\begin{eqnarray}
\hat f(x)=\frac{2m}{\hbar^2}\, V(x),\label{f}
\hspace{2cm}\epsilon=\frac{2m}{\hbar^2}\, E.
\end{eqnarray}
As mentioned before, we use the trigonometric basis set to find the
energy spectrum. That is, since we need to choose a finite subspace
of a countably infinite basis, we restrict ourselves to the finite
region $-L<x<L$. This means that we can expand the solution as
\begin{eqnarray}
\psi(x)=\sum_{i=1}^{2} \sum_{m=0}^{\infty} A_{m,i} \,\,\,
g_i\left(\frac{m \pi x}{L}\right), \label{eqpsitrigonometric}
\end{eqnarray}
where
\begin{eqnarray}
\left\{
  \begin{array}{ll}
  g_1\left(\frac{m \pi
x}{L}\right)=\frac{1}{\sqrt{LR_m}} \sin\left(\frac{m \pi
x}{L}\right), &  \\
  g_2\left(\frac{m \pi
x}{L}\right)=\frac{1}{\sqrt{LR_m}}\cos\left(\frac{m \pi
x}{L}\right), & \\
  \end{array}\hspace{1cm}R_{m}=\left\{
  \begin{array}{ll}
    2 & \hspace{1cm}\hbox{m=0,} \\
    1 & \hspace {1cm}\hbox{otherwise.}
  \end{array}\right.
\right.
\end{eqnarray}
In the above choice of the basis, we are implicitly assuming periodic
boundary condition. We can also make the following expansion
\begin{eqnarray}
\hat f(x) \psi(x)=\sum_i \sum_{m} B_{m,i} \,\,\, g_i\left(\frac{m
\pi x}{L}\right),\label{eqV}
\end{eqnarray}
where $B_{m,i}$ are coefficients that can be determined once $\hat
f(x)$ is specified. By substituting Eqs.~(\ref{eqpsitrigonometric})
and (\ref{eqV}) into Eq.~(\ref{ODE}) we obtain
\begin{eqnarray}
\sum_{m,i}\left[\left(\frac{m \pi}{L}\right)^2
A_{m,i}+B_{m,i}\right]g_i\left(\frac{m \pi
x}{L}\right)=\epsilon\sum_{m,i}A_{m,i}\,g_i\left(\frac{m \pi
x}{L}\right).\label{eqAB1}
\end{eqnarray}
Because of the linear independence of $g_i(\frac{m \pi x}{L})$,
every term in the summation must satisfy
\begin{eqnarray}
\left(\frac{m \pi}{L}\right)^2 A_{m,i}+B_{m,i}=\epsilon\,
A_{m,i}.\label{eqAB2}
\end{eqnarray}
It only remains to determine the matrix $B$. Using Eqs.~(\ref{eqV})
and (\ref{eqpsitrigonometric}) we have
\begin{eqnarray}
\sum_{m,i} B_{m,i} g_i\left(\frac{m \pi
x}{L}\right)\,\,\,=\sum_{m,i} A_{m,i} \hat f(x) g_i\left(\frac{m
\pi x}{L}\right).
\end{eqnarray}
By multiplying both sides of the above equation by $g_{i'}(\frac{m'
\pi x}{L})$ and integrating over the $x$-space and using the
orthonormality condition of the basis functions, one finds
\begin{eqnarray}\label{eqbmi}
B_{m,i}=\sum_{m',i'} A_{m',i'} \int_{-L}^{L} g_i\left(\frac{m \pi
x}{L}\right)\,\,\,\hat f(x)\,\,\, g_{i'}\left(\frac{m' \pi
x}{L}\right)\,\,\, dx= \sum_{m',i'} A_{m',i'} C_{m,m',i,i'}.
\end{eqnarray}
Therefore we can rewrite Eq.\ (\ref{eqAB2}) as
\begin{eqnarray}
\left(\frac{m \pi}{L}\right)^2 A_{m,i}+ \sum_{m',i'}
C_{m,m',i,i'}\,\, A_{m',i'}=\epsilon\, A_{m,i}.\label{eqAC}
\end{eqnarray}
Where the coefficients $C_{m,m',i,i'}$ are defined by Eq.
(\ref{eqbmi}). It is obvious that the presence of the operator $\hat
f(x)$ in Eq.\ (\ref{ODE}), leads to nonzero coefficients
$C_{m,m',i,i'}$ in Eq.\ (\ref{eqAC}), which in principle could
couple all of the matrix elements of $A$. It is easy to see that the
more basis functions we include, the closer our solution will be to
the exact one. By selecting a finite subset of the basis functions,
{\it e.g.} choosing the first $2N$ which could be accomplished by
letting the index $m$ run from 1 to $N$ in the summations,
Eq.~(\ref{eqAC}) can be written as
\begin{eqnarray}
D\, A=\epsilon\, A, \label{eqmatrix}
\end{eqnarray}
where $D$ is a square matrix with $(2N) \times (2N)$ elements. The
eigenvalues and eigenfunctions of the Schr\"{o}dinger equation are
approximately equal to the corresponding quantities of the matrix
$D$. That is the solution to this matrix equation simultaneously
yields $2N$ sought after eigenstates and eigenvalues. We are free to
adjust two parameters: $2N$, the number of basis elements used and
the length of the spatial region, $2L$. This length should be
preferably larger than spatial spreading of all the sought after
wave functions. However, if $2L$ is chosen to be too large we loose
overall accuracy. It is important to note that for each $N$, $L$ has
to be properly adjusted. This is in fact the  optimization procedure
and we denote this optimal quantity by $\hat{L}(N)$: For a few
values of $N$ we compute $\epsilon(N,L)$ which invariably has an
inflection point in the periodic boundary condition. Therefore, all
we have to do is to compute the position of this inflection point
and compute an interpolating function for obtaining $\hat{L}(N)$.

\section{Applications}
In this section, for illustrative purposes, we first apply the
optimization procedure to find the bound states of a Simple Harmonic
Oscillator (SHO) which is an exactly solvable case. We then apply
this method to two perturbed harmonic oscillators, the first with a
quartic anharmonic term, and the second with a rapidly oscillating
trigonometric anharmonic term.

\subsection{Simple Harmonic Oscillator}
The Schr\"{o}dinger equation for SHO is
\begin{equation}\label{SHO}
-\frac{\hbar^2}{2m}\frac{d^2\psi(x)}{dx^2}+\frac{1}{2}m \omega^2
x^2 \psi(x)=E\psi(x),
\end{equation}
where $\omega$ is the natural frequency of the oscillator. The
dimensionless equation now reads
\begin{equation}\label{SHO2}
-\frac{d^2\psi(x)}{dx^2}+x^2 \psi(x)=\epsilon\psi(x),\hspace{0.5cm}
\mbox{where} \hspace{1cm} \epsilon=\frac{2E}{\hbar\omega}.
\end{equation}
The exact solutions are
\begin{eqnarray}
\psi_n(x)=\left(\frac{\omega}{\pi}\right)^{1/4}\frac{H_n(
\sqrt{\omega}x)} {\sqrt{2^n n!}}e^{-\omega
x^2/2},\hspace{1cm}\epsilon_n=\left(n+\frac{1}{2}\right)\hbar\omega,\hspace{1cm}n=0,1,2,...,\label{SHO1}
\end{eqnarray}
where $H_n(x)$ denote the Hermite polynomials.

\begin{figure}
 \centering
\includegraphics[width=8cm]{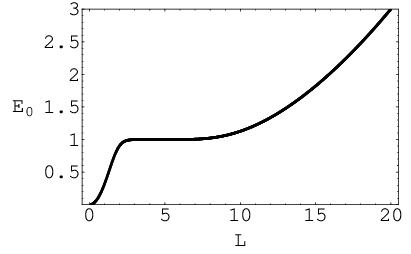}
\caption{Ground state energy  for SHO versus $L$ for $N=5$, using SM
in units where $\hbar\omega=2$.}\label{figNL}
\end{figure}

\begin{figure}
  \centering
  \includegraphics[width=10cm]{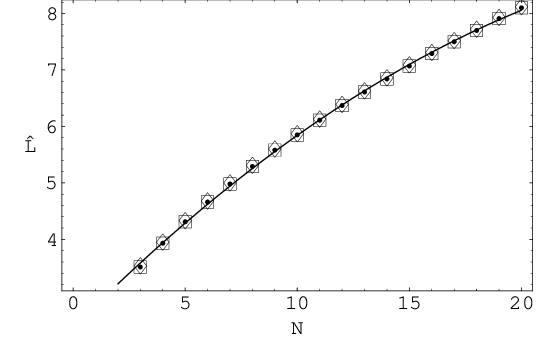}\\
  \caption{$\hat{L}$ versus $N$ and its interpolating function.}\label{figLN2}
\end{figure}

\begin{figure}
\centering
\includegraphics[width=10cm]{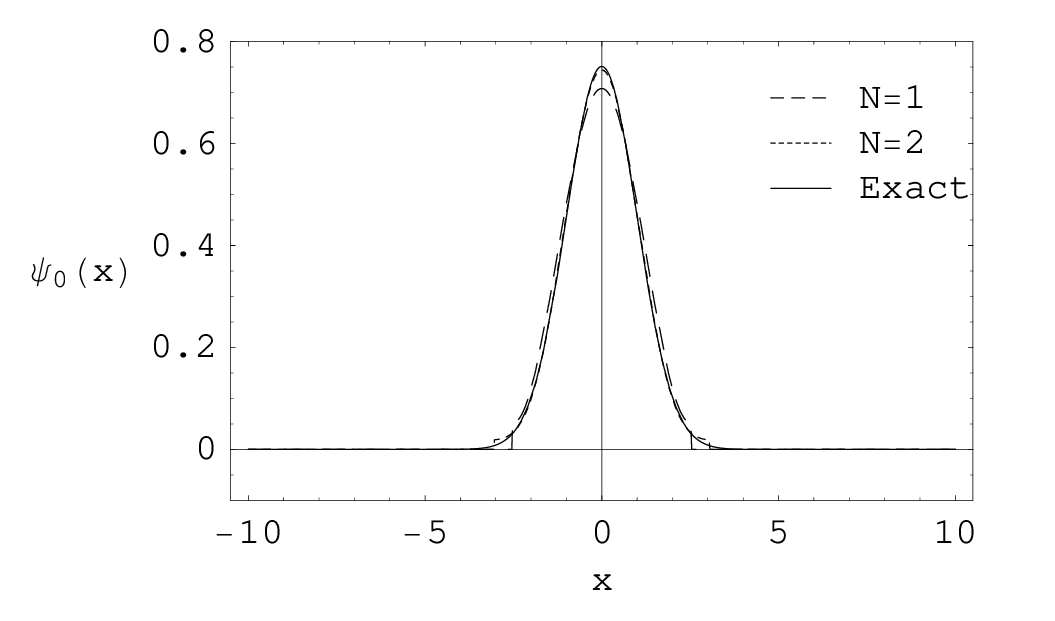}
\caption{The exact and approximate ground state wave functions of
SHO for $N=\{1,2\}$ with optimized $\hat{L}=\{2.52479,3.04635\}$,
respectively. } \label{fig SHO}
\end{figure}

In Fig.~\ref{figNL} we showed the ground state energy computed for
$N=5$ as a function of $L$ using periodic boundary condition. Note
the existence of the inflection point that determines $\hat{L}(5)$.
We repeat this procedure for a few other values of $N$. After
plotting these values we can obtain an interpolating function
$\hat{L}(N)$. In Fig. \ref{figLN2} we show our results for
$\hat{L}(N)$ and its interpolating function. Having determined
$\hat{L}(N)$, we can proceed to compute the bound states. Table
\ref{Table Oscillator} shows the results for the first 10
energy eigenvalues for $N=100$.  The left part of Figure \ref{fig SHO}
shows the exact and approximate ground state wave functions for
$N=\{3,5,7\}$ with a fixed un-optimized, namely, $L=10$. The right part of the
same figure shows the exact and approximate ground state wave
functions for $N=\{1,2\}$
 with optimized $\hat{L}=\{2.52479,3.04635\}$,
respectively.

\begin{table}
  \centering
\begin{tabular}{c|c|c}
  $n$   & $\epsilon_n^{exact}$        & $\mathrm{error}$   \\ \hline
   0    &1       & $2.6\times10^{-139}$                \\ \hline
   1    &3       & $1.1\times10^{-133}$                \\ \hline
   2    &5       & $5.9\times10^{-134}$                \\ \hline
   3    &7       & $7.5\times10^{-129}$                \\ \hline
   4    &9       & $2.2\times10^{-129}$                \\ \hline
   5    &11      & $1.5\times10^{-124}$                \\ \hline
   6    &13      & $3.1\times10^{-125}$                \\ \hline
   7    &15      & $1.3\times10^{-120}$                \\ \hline
   8    &17      & $2.4\times10^{-121}$                \\ \hline
   9    &19      & $7.1\times10^{-117}$                \\
\end{tabular}
  \caption{The results for the first 10 eigenvalues and eigenfunctions of the SHO with $N=100$. }
  \label{Table Oscillator}
\end{table}

%*************************************Anharmonic Oscillator******************
\subsection{Anharmonic Oscillator with a quartic term}
Now we apply this method to an anharmonic oscillator which has a
quartic term. The Schr\"{o}dinger equation for this not exactly
solvable model is given by
\begin{equation}\label{ASHO}
-\frac{\hbar^2}{2m}\frac{d^2\psi(x)}{dx^2}+\left(\frac{1}{2}m\omega^2
x^2+\gamma x^4\right) \psi(x)=E\psi(x).
\end{equation}
The results that we have obtained using $N=100$ are extremely
accurate (see Table \ref{Table anharmonic}).  This problem is also
approximately solved using the zero, first and second order
variational sturmian approximation \cite{mostafazadeh}. Moreover, in
Ref.~\cite{exactanhrmonic} the highly accurate results are obtained
with 90 significant digits.

\begin{table}
\centering
\begin{tabular}{c|c|c}
$n$&$\epsilon_n$   & $SD$  \\ \hline
0&1.0652855095437176888570916287890930843044864178189& 124        \\ \hline
1&3.3068720131529135071281216846928690495946552097516& 121   \\ \hline
2&5.7479592688335633047335031184771312788809760663913& 120           \\ \hline
3&8.3526778257857547121552577346436977053951052605059& 118   \\ \hline
4&11.098595622633043011086458749297403250621831282348& 118  \\ \hline
5&13.969926197742799300973433956842133961140713634295& 116  \\ \hline
6&16.954794686144151337692616508817134375549987258361& 114  \\ \hline
7&20.043863604188461233641421107385111570572266905826& 115   \\ \hline
8&23.229552179939289070647087434323318243534938599487& 112        \\ \hline
9&26.505554752536617417469503006738723676057932189542& 110  \\
\end{tabular}
\caption{The first ten energy levels of the anharmonic oscillator
whose dimensionless form is $(-d^2/dx^2+x^2+\bar\gamma
x^4)\psi(x)=\epsilon\psi(x)$ and  $\bar\gamma=4 \gamma/(m
\omega^4)=0.1$. We used $N=100$ basis functions and  $SD$ denotes
the number of significant digits.}\label{Table anharmonic}
\end{table}
%*************************************Oscillatory****************************

\subsection{Harmonic Oscillator perturbed by a rapid oscillation}
\begin{figure}
\centering
\begin{tabular}{ccc} \includegraphics[width=9cm]{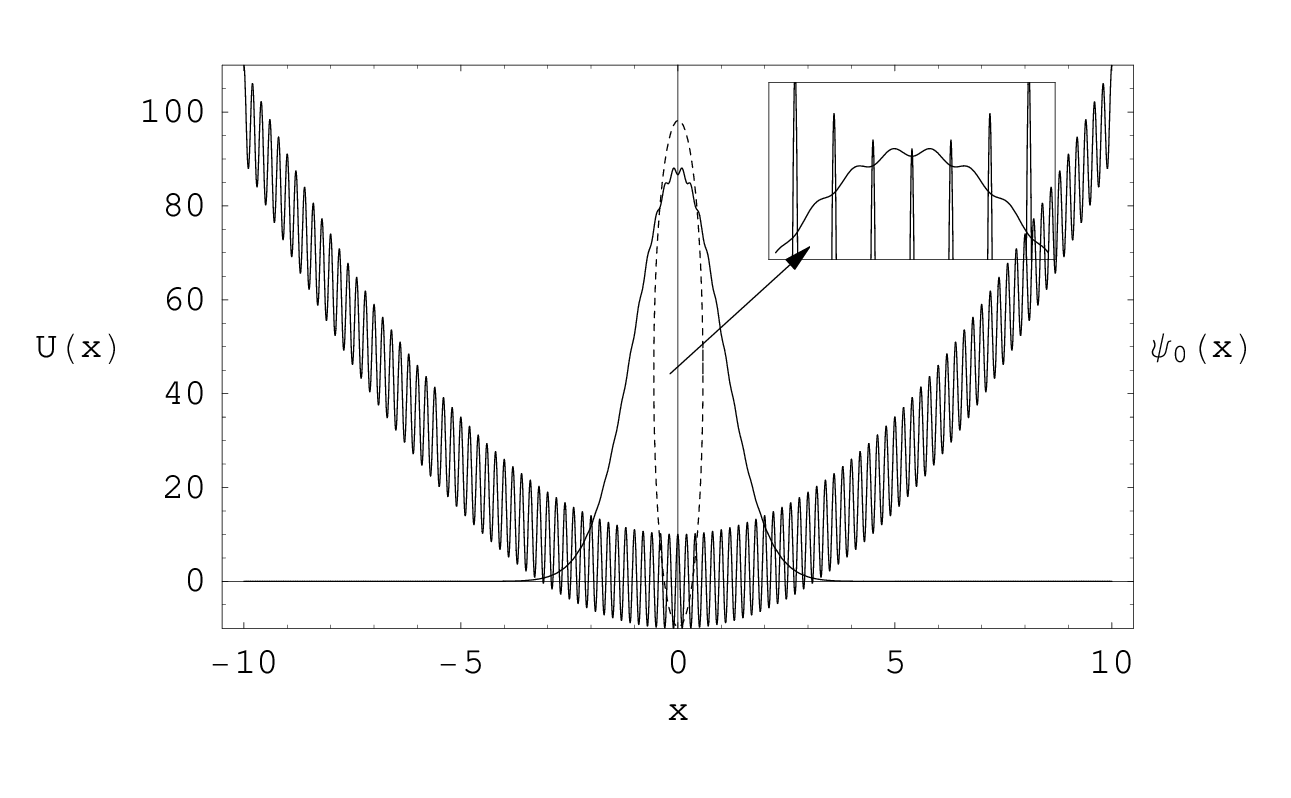}
&\hspace{1.cm}& \includegraphics[width=7cm]{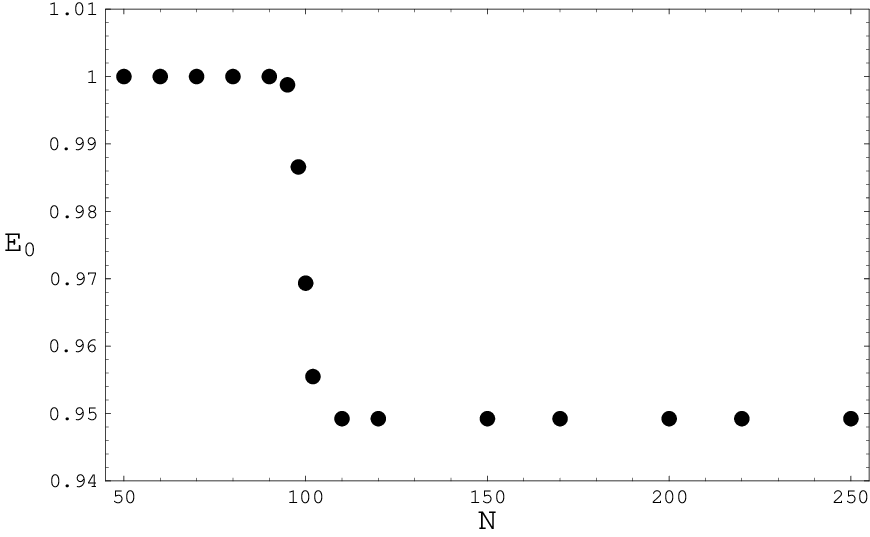}
\end{tabular}
\caption{Left, the potential of the harmonic oscillator perturbed by
rapid oscillations, whose dimensionless Schr\"{o}dinger equation
$(-d^2/dx^2+x^2+\alpha' \cos(\beta'\pi x))\psi(x)=E'\psi(x)$. We
have chosen the parameters $\beta'=\sqrt{2/m} \beta/\omega=10$, and
$\alpha'=(2/\hbar\omega)\alpha=10$. Superimposed on the same graph
is the ground state wave function calculated with $N=150$. Right,
the ground state energy versus $N$.} \label{PHO}
\end{figure}

A rather interesting example is the harmonic oscillator perturbed by
a rapid oscillation whose  Schr\"{o}dinger equation is given by
\begin{equation}\label{SHOR}
-\frac{\hbar^2}{2m}\frac{d^2\psi(x)}{dx^2}+\left(\frac{1}{2}m
\omega^2 x^2+\alpha \cos(\beta \pi x)\right) \psi(x)=E\psi(x),
\end{equation}
where $\omega$ is the natural frequency of the oscillator and
$\alpha$ and $\beta$ are arbitrary constants.  This differential
equation is not exactly solvable and for large $\beta$ the behavior
of the potential is very oscillatory and centered around the curve
$\frac{1}{2} m\omega x^2$.  The results for the ground state are
shown in the Fig.~\ref{PHO}. In the left part of the figure, we
showed the full potential, the ground state wave function, and a
zoomed box highlighting the fine structural behavior of the wave
function. In the right part of the figure, we showed the ground state
energy $E_0$ versus $N$. Note that for $N$ smaller than $\beta
\hat{L}$ (100 here) this method is not sensitive enough to respond
to the rapidly oscillating part of the potential and the results are
very close to those of the (unperturbed) SHO. As is apparent from
the figure, for $N$ slightly larger than $\beta \hat{L}$, the energy
eigenvalue approaches the exact energy eigenvalue as $N$ increases.

It is now worth mentioning the two main advantages of this technique
with respect to Ref.~\cite{p4}: First, the method of Ref.~\cite{p4}
is only applicable for the bounded power low potentials, but this
method works for the general class of the bounded $C^\infty$
potentials. Second, in our method the potentials does not have to to
be symmetric. Note that we can obtain arbitrary accuracy by
increasing the number of the basis functions. But, the speed of the
calculation decreases due to the presence of large matrices. Indeed,
one of the time consuming parts of the algorithm is finding the
coefficients $C_{m,m',i,i'}$ that are defined by Eq.~(\ref{eqbmi}).
One advantage of our method with respect to other spectral methods
such as Chebyshev spectral method is that these coefficients can be
obtained analytically before numerical diagonalization of the
Hamiltonian for a large class of potentials.

\section{Conclusions}
We have used the optimized trigonometric finite basis method as an
extremely accurate technique for obtaining energy eigenvalues and
eigenfunctions of the bound states of the time-independent
Schr\"{o}dinger equation. The optimization procedure is based on the
presence of an inflection point in the plot of the energy eigenvalue
versus the domain of the basis. We applied this method to the
quartic anharmonic oscillator case which is not exactly solvable and
found the solutions with high accuracy. Also, we solved the problem
of SHO perturbed by a trigonometric anharmonic term and showed that
how the optimization scheme properly handles problems involving
potentials with rapid oscillations.

\end{document}